\documentclass[%
superscriptaddress,
 amsmath,amssymb,
 aps,
twocolumn,
]{revtex4-1}

\usepackage{graphicx}
\usepackage{dcolumn}
\usepackage{bm}

\usepackage{color}

\begin{document}
\title{Observation of optical absorption correlated with surface state of topological insulator}

\author{Jiwon Jeon}
\affiliation{Department of Physics, University of Seoul, Seoul 130-743, Korea}

\author{Kwangnam Yu}
\affiliation{Department of Physics, University of Seoul, Seoul 130-743, Korea}

\author{Jiho Kim}
\affiliation{Department of Physics, University of Seoul, Seoul 130-743, Korea}

\author{Jisoo Moon}
\affiliation{Department of Physics and Astronomy, Rutgers, the State University of New Jersey, Piscataway, New Jersey 08854, USA}

\author{Seongshik Oh}
\affiliation{Department of Physics and Astronomy, Rutgers, the State University of New Jersey, Piscataway, New Jersey 08854, USA}

\author{E. J. Choi}
\email{echoi@uos.ac.kr}
\affiliation{Department of Physics, University of Seoul, Seoul 130-743, Korea}

\date{\today}

\begin{abstract}
We performed broadband optical transmission measurements of Bi$_{2}$Se$_{3}$ and In-doped
$\rm{(Bi_{1-x}In_{x})_{2}Se_{3}}$ thin films, where in the latter the spin-orbit coupling (SOC) strength 
can be tuned by introducing In. Drude and interband transitions exhibit In-dependent changes that are consistent with evolution from metallic (x=0) to insulating (x=1) nature of the end compounds. Most notably,
an optical absorption peak located at $\hbar\omega$=1eV in Bi$_{2}$Se$_{3}$
is completely quenched at x=0.06, the critical concentration where the phase transition from TI into non-TI takes place. For this x, the surface state (SS) is vanished from the band structure as well. The correlation between
the 1eV optical peak and the SS in the x-dependences suggests that the peak is associated with the SS.
We further show that  when Bi$_{2}$Se$_{3}$ is electrically gated, the 1eV-peak becomes stronger(weaker) 
when electron is depleted from (accumulated into) the SS.  
These observations combined together demonstrate that under the $\hbar\omega$=1eV illumination 
electron is excited from a bulk band into the topological surface band of Bi$_{2}$Se$_{3}$. 
The optical population of surface band is of significant importance not only for fundamental study but also 
for TI-based optoelectronic device application.
\end{abstract}

\pacs{
68.65.Pq 
78.30.-j 
78.67.Wj 
}
\maketitle

{\it Introduction.}
Topological insulator (TI) is a novel state of matter characterized by insulating bulk and metallic surface\cite{fu2007topological,xia2009observation,hasan2010colloquium,moore2010birth,qi2011topological}.
The surface state, topologically protected and chirally-textured, 
supports dissipationless spin-conserving current, applicable  
for quantum devices\cite{roushan2009topological,hsieh2009tunable,li2014electrical}.
Optically, various kinds of electron transitions occur in a TI under photo illumination both in the surface and in the bulk as demonstrated by numerous previous experiments:
For the archetypal TI material Bi$_{2}$Se$_{3}$,  
intraband (Drude) transition and Kerr rotation of the surface carrier were observed in THz measurement  \cite{aguilar2012terahertz}.   
In the infrared range, Post {\it et al} measured the interband transition from
bulk valence-band (VB) to bulk conduction-band (CB), VB$\rightarrow$CB, and determined upper bound of Drude weight of SS \cite{post2015sum}.
Also Falsetti {\it et al} observed the infrared Berreman resonance of the surface electron in 
 Bi$_{2}$Se$_{3}$ thin films \cite{falsetti2018infrared}.
On the other hand, for periodically modulated Bi$_{2}$Se$_{3}$, plasmonic excitation of the surface electron  \cite{di2013observation} and 
the plasmon-phonon interaction \cite{in2018control} were observed at THz frequencies.

One interesting optical absorption that TI can host yet has not been detected is an excitation of electron from bulk band into SS. This particular transition between bulk and surface, VB$\rightarrow$SS, will provide rare opportunity to study how 
the surface and bulk are connected optically. Also, when the surface electron is populated by this optical transition,  
the surface electron is increased and therefore           
the topological current ($\propto$ surface electron) is enhanced, which can boost 
the performance of TI as optoelectronic device such as   photogalvanics  and optical imaging display 
\cite{mciver2012control, yue2017nanometric}. 
In fact Sobota {\it et al} showed that  a similar transition 
can occur from bulk CB to 2nd SS, CB$\rightarrow$2nd SS, at 
a visible frequency  \cite{sobota2013direct}. (Here the 2nd SS refers to another SS that lies above the 1st  
or fundamental SS). However the optical transition into the 1st SS or fundamental SS, VB$\rightarrow$SS, was not 
reported yet.
It is not clear at this point whether
the lack of the SS-populating  optical transition  
is due to that other transition such as VB$\rightarrow$CB  is overwhelmingly stronger,
making the detection difficult \cite{li2015optical}, or more fundamentally,  this particular transition
is forbidden by optical selection rule.

Here we performed broadband optical absorption measurement of Bi$_{2}$Se$_{3}$ and $\rm{(Bi_{1-x}In_{x})_{2}Se_{3}}$ thin films from far-IR to UV frequencies. 
In $\rm{(Bi_{1-x}In_{x})_{2}Se_{3}}$, the spin-orbit coupling (SOC) strength is modulated by means of 
the indium (In) substitution. As In-content is increased the SOC of Bi$_{2}$Se$_{3}$ is decreased 
and consequently the topological property is softened. Eventually, 
at certain substitution level, the topological SS is completely quenched from the band structure. Accordingly  any SS-related optical transition will be removed from the wide-range optical excitation spectrum, which
in fact  offers us an invaluable means to find the SS-population transition in particular.
Our measurement shows signatures that such optical absorption may exist.
\newline

{\it Experiment.}
High quality epitaxial  Bi$_{2}$Se$_{3}$  and $\rm(Bi_{1-x}In_{x})_{2}Se_{3}$ thin films were grown on
Al$_{2}$O$_{3}$ and SiO$_{2}$/Si substrates using the 
MBE method \cite{bansal2011epitaxial}. Optical transmittance T($\omega$)
was measured from Far-infrared to UV by using Fourier transform infrared spectroscopy (FTIR) spectrometer
in combination with spectroscopic ellipsometer. For gate-dependent optical measurement 
gate-voltage $V_{\rm{G}}$ was applied between $\rm{Bi_{2}Se_{3}}$ and Si of the substrate.
Optical conductivity  $\sigma_{1}(\omega)$ was calculated from transmission data  
through rigorous Kramers-Kronig transformation by using RefFit\cite{kuzmenko2005kramers}. The experimental details  are described in Supplemental Material 1\cite{suppley} and references therein. 
\newline

{\it Results.}
Figure 1 shows the wide-range optical conductivity  $\sigma_{1}(\omega)$ of 
the 50QL-thick  $\rm{Bi_{2}Se_{3}}$ film. 
In the Far-infrared region, $\sigma_{1}(\omega)$ consists of Drude absorption and optical phonon peak, both
coming from the bulk, where the former one arises from the Se-vacancy driven carrier \cite{post2013thickness,di2012optical}.  
For $\hbar\omega$$>$0.25eV, the interband (IB) transition VB$\rightarrow$CB leads to 
the rapid rise of $\sigma_{1}(\omega)$. 
Note that there is an absorption peak at $\hbar\omega$=1eV   
($\equiv$Peak-A hereafter) which we will pay particular attention to.
\newline

In Figure 2 we show optical conductivity measured for a series of In-substituted  $\rm{(Bi_{1-x}In_{x})_{2}Se_{3}}$ films. The In-concentration $x$ was varied for 0$\leq$x$\leq$0.9 range. 
Previous studies showed that as Bi is replaced by the light element In, the spin-orbit interaction  
is reduced and the topological property of Bi$_{2}$Se$_{3}$ becomes weaker \cite{brahlek2012topological, salehi2016finite, wu2013sudden, sim2015tunable}. 
At a critical concentration   x$_{c}$, 
phase transition from TI to non-TI (NTI) phase occurs where the bulk band gap is closed, and CB and VB begin to reinvert.  The x$_{c}$ lies between x=0.04 and x=0.06 
depending on the film thickness, and for  at x$\geq$x$_{c}$  the topological SS is completely vanished \cite{brahlek2012topological,salehi2016finite}. The
$\sigma_{1}(\omega)$ shows that Peak-A becomes weaker  
as $x$ increases. For quantitative analysis of this behavior
we isolate Peak-A by removing background conductivity from $\sigma_{1}(\omega)$ as 
$\sigma_{1}^{A}(\omega)$=$\sigma_{1}(\omega)$-$\sigma_{1}^{BG}(\omega)$ 
as illustrated in the inset of Figure 2d, (a polynomial function was used for the $\sigma_{1}^{BG}$)
and calculated the strength of Peak-A as $S$=$\int\sigma_{1}^{A}(\omega)d\omega$.
Figure 2d 
shows that $S$ is quenched at x=0.06.  To double check this behavior 
we performed independent analysis of Peak-A:  
we calculate the second derivative $\frac{d^{2}\sigma_{1}}{dE^{2}}$   
 and measure the distance (w) and depth ($d$) of the extrema pattern,  which allows
determination of strength $S$ (=$\frac{1}{12}$$\sqrt{\frac{\pi}{6}}$$\cdot$dw$^{3}$) as well as width (= $\frac{1}{\sqrt{3}}$$\cdot$w) and  height (=$\frac{1}{12}$ $\cdot$dw$^{2}$) of Peak-A. (See Supplemental Fig. S1 for details\cite{suppley}). 
The $S$ is quenched at x=0.06  again, which confirms the $S$=$\int\sigma_{1}^{A}(\omega)d\omega$ analysis.
Importantly x=0.06 
is the critical x$_{c}$ for the TI$\rightarrow$NTI transition for the thickness $d$=50QL of our films: that is, the SS is vanished at this x. This correlation of Peak-A with the TI$\rightarrow$NTI  transition
strongly suggests that Peak-A is related with the topological SS of Bi$_{2}$Se$_{3}$. We emphasize that this behavior is strikingly different from those of the other optical absorption features: 
In Supplemental Fig.S2\cite{suppley}, we show that the Drude absorption is 
vanished at x$\sim$0.5, the phonon peak splits at x= 0.12, and the IB survives up to x=0.9. 
(For x=1, $\rm{In_{2}Se_{3}}$ is a large gap band insulator with $E_{\rm{g}}$ $>$ 1.5eV) 
Note that none of these features are correlated with x$_{c}$. In contrast     
Peak-A manifests clear correlation with x$_{c}$ and is the only absorption of such kind. 
\newline

Given the correlation of Peak-A with SS, one can propose possible pictures on how Peak-A is created.  Specifically, Peak-A can arise when (1) SS electron is excited into 
empty state lying 1eV above, or alternatively
(2) electron lying at 1eV below is excited into the SS. In both scenarios
 Peak-A becomes extinct when SS is suppressed at x$_{c}$.  
To find out which scenario is correct, we performed electrical gating experiment on the Bi$_{2}$Se$_{3}$ film ($d$=8QL). 
For this a Bi$_{2}$Se$_{3}$ film was grown on SiO$_{2}$/Si substrate and optical transmission was measured while gate voltage  $V_{\rm{G}}$ is applied between the film and Si. 
In this back-gate configuration, the Fermi energy E$_{F}$ shifts down (up) 
for the negative (positive) $V_{\rm{G}}$ due to electron depletion (accumulation) in the film.
Fig.3(b) shows that T($V_{\rm{G}}$)/T(0) changes in the Far-IR, mid-IR, and at 1eV.
Figure 3(c) shows that 
Peak-A becomes stronger (weaker) for negative (positive) $V_{\rm{G}}$. Such change  
supports the scenario (2) over (1) for the following reason: For  $V_{\rm{G}}$$<$0  the electron occupation of SS is 
reduced and more empty SS become available, which strengthens the transition of (2), which 
agrees with 
the increase of Peak-A strength. This relation is visualized in Fig.4.  In the scenario (1), on the other hand, the surface electron is decreased and 
the peak becomes weaker, opposite to the observed behavior of Peak-A. 
Therefore, the  $V_{\rm{G}}$-dependent result demonstrates that Peak-A arises most likely by excitation of electron from a state lying 1eV below into the SS. In this transition electron occupation of SS is increased, or equally,
the SS is optically populated by illuminating Bi$_{2}$Se$_{3}$ with $\hbar\omega$=1eV. 
Here we remark that 
 the $V_{\rm{G}}$-dependent change in Figure 3c is very small, less than even 0.1\%. 
Nevertheless the Peak-A change is successfully measured, demonstrating the superior sensitivity and stability of our experiment. 
For later analysis
we calculate the $V_{\rm{G}}$-dependent $\sigma_{1}(\omega)$ from the T($V_{\rm{G}}$)/T(0) data 
\cite{yu2016infrared, yu2019infrared,yu2019gate} and show it in Figure 3(d). 
\newline

With the nature  of Peak-A been identified, the next question to be addresed is the origin of the initial state. Before we discuss
this issue, we give further thoughts on the gate-dependent growth of Peak-A. Figure 3 implies that 
the SS-population will become stronger if E$_{F}$ could be brought down further. The latter would be possible when  $V_{\rm{G}}$ is applied to high value beyond the limit of our measurement, where such high-gating was in fact demonstrated experimentally \cite{bansal2014robust}.  Here we will consider how
large Peak-A will grow in the strong-gating regime.  
In Figure 4a, we show the bulk interband transition in the mid-IR range.  
The onset energy  ($\equiv$E$_{op}$) of this transition corresponds to the thick arrow in Fig.4(b). 
The E$_{op}$ increases when the Fermi level shifts up. Fig.4(c) shows that 
the increase rate is dE$_{op}$/d$V_{\rm{G}}$ =1.74$\times10^{-4}$[eV/V]$\equiv$$a$. 
In the mean time 
the $S$=$\int\sigma_{1}^{A}(\omega)d\omega$ calculated from Figure3(d) decreases 
at the rate 
d$s$/d$V_{\rm{G}}$ =-6.06$\times10^{-4}$[1/V]$\equiv$$b$, where $s$=S/S(0V) is the normalized $S$ by 
ungated $S(0)$. 
Given $a$ and $b$ 
we can eliminate $V_{\rm{G}}$ and obtain the S-change against E$_{F}$ as d$s$/dE$_{F}$ = (d$s$/d$V_{\rm{G}}$)$\cdot$(d$V_{\rm{G}}$/dE$_{F}$)= $b$/($a$/2)=6.96[1/eV].
Here d$V_{\rm{G}}$/dE$_{F}$=$(a/2)^{-1}$  was derived by utilizing the Bernstein-Moss relation\cite{burstein1954anomalous,moss1954interpretation,hamberg1984band}, 
namely, dE$_{F}$/d$V_{G}$= $\frac{1}{2}$$\times$dE$_{op}$/d$V_{\rm{G}}$=$\frac{a}{2}$ where   
the factor $\frac{1}{2}$ comes from $\frac{1}{m_{CV}^*}=\frac{1}{m_{CB}^*}+\frac{1}{m_{VB}^*}$ 
and $m_{CB}^*$=$m_{VB}^*$\cite{analytis2010bulk}.
This result d$s$/dE$_{F}$=6.96[1/eV] enables us estimate $S$ at high gating:
For the pristine, electron-doped Bi$_{2}$Se$_{3}$ films like ours, the Fermi level lies typically at 
$E_{F}$$\sim$0.1eV from the CB bottom (CBB).  
If the gating shifts $E_{F}$ down to the CBB, i.e, $\Delta$E$_{F}$=0.1, then
$s$ will increase approximately by $\Delta$$s$ $\approx$ [$\frac{ds}{d\rm{E}_{F}}$]$\cdot$ $\Delta$E$_{F}$=0.69.
that is, Peak-A grows by $\sim$70$\%$ compared with the ungated strength. 
If E$_{F}$ is shifted further to the Dirac point, the latter lying $\sim$0.2eV from the CBB, we have 
$s$=$s(0)$+$\Delta$$s$$\approx$1+[$\frac{ds}{d\rm{E}_{F}}$]$\cdot$(0.1+0.2)=3.09. 
That is, Peak-A grows as large as three-times.   (Here we assumed
 $\frac{ds}{d\rm{E}_{F}}$ is constant for simplicity, neglecting its $E_{F}$-dependence.)
This estimation shows that substantial increase of the peak will occur at high-gating.
From this exercise we also learn that if $S$ is precisely characterized as function of E$_{F}$, 
it could be used to determine the location of the Fermi level in Bi$_{2}$Se$_{3}$ films, whereas
usually more difficult ARPES should be performed.
\newline

{\it Discussion.}
We now search for possible candidate for the initial state of the Peak-A. For this we refer to the
band structure of Bi$_{2}$Se$_{3}$ reported in experimental 
\cite{hsieh2009tunable,nechaev2013evidence,hasan2010colloquium,pan2011electronic,wray2010observation,dubroka2017interband,piot2016hole} and theoretical \cite{aguilera2013g,guo2016tuning,forster2015ab,hermanowicz2019iodine} literatures, and 
schematically redrew it in Figure5. 
In Figure 5 note that  there is an energy branch lying  
$\sim$1eV below the SS.  Interestingly, 
this branch E(k) runs in near-parallel with the SS. 
If we consider the optical transition from this E(k) branch (=i) to SS (=f), their parallel dispersion
$\nabla_{k}\rm{E}^{i}(k)$$\cong$$\nabla_{k}\rm{E}^{f}(k)$  
leads to strong absorption due to that transition 
strength    
S$\sim$$\int \frac{M_{fi}}{|\nabla_{k}\rm{E}^{f}(k)-\nabla_{\emph{k}}\rm{E}^{i}|} d^{2}k$ 
becomes divergent. This yields a pronounced absorption at $\hbar\omega$=E$_{f}$-E$_{i}$=1eV, which 
agrees with the profile of Peak-A.  
(Here the transition matrix element M$_{fi}$ is assumed to be constant for simplicity.)
Therefore this 1eV-E(k) is a plausible candidate for the i-state.  
We think that this assignment can be confirmed when theoretical calculation of $\sigma_{1}(\omega)$, not available currently, is performed. We remark that optical transition of  Bi$_{2}$Se$_{3}$ 
between bulk and surface in particular was poorly 
studied so far with a rare exception\cite{li2015optical}. 
To consider another candidate native defect such as Se-vacancy can produce defect levels below  $E_{F}$.
However their energy locations are not well known, and generally such localized, dispersionless levels do not
fulfill the 
$\nabla_{k}\rm{E}^{i}(k)$$\cong$$\nabla_{k}\rm{E}^{f}(k)$  condition. 
We emphasize that, while supporting works should follow to definitely identify the 1eV-bulk E(k) as origin of i-state, the occurrence of the optical population of SS in Bi$_{2}$Se$_{3}$ is evident 
judging from the properties of Peak-A we have unveiled regardless of the i-state origin.

To make further remark on the T($V_{\rm{G}}$)/T(0) data,  Fig.3(b) shows that gate-dependent change occurs in the Far-IR and mid-IR ranges also. Similar change was reported for bulk-insulating $\rm{(Bi_{1-x}Sb_{x})_{2}Te_{3}}$ films\cite{whitney2016gate}.
While for $\rm{(Bi_{1-x}Sb_{x})_{2}Te_{3}}$ the mid-IR modulation peaks at $\sim$0.3eV, the modulation for Bi$_{2}$Se$_{3}$ occurs at higher energy, peaked at 0.45 eV. This difference is  attributed to that the interband transition  taking  place at E$_{op}$=E$_{g}$+2E$_{F}$ is higher for Bi$_{2}$Se$_{3}$ where  E$_{F}$ is significant ($\sim$0.1 eV) compared with  the insulating $\rm{(Bi_{1-x}Sb_{x})_{2}Te_{3}}$  where  E$_{F}$  is considered to be much lower.    
Also, while the modulations in mid-IR and far-IR inevitably contain contribution from both bulk and surface states, the modulation strength of the 1eV feature is weaker,  which may further support the surface-related origin. Further quantitative  analysis and comparison will be published separately.
\newline


In conclusion we performed broadband optical absorption measurement on pristine Bi$_{2}$Se$_{3}$ 
and In-substituted $\rm{(Bi_{1-x}In_{x})_{2}Se_{3}}$ thin films, as well as electrically gated  Bi$_{2}$Se$_{3}$. The absorption Peak-A that occurs at $\hbar\omega$=1eV 
showed clear correlation with the In-driven TI-NTI phase transition: it is activated at x$<$x$_{c}$ (TI-phase)
but is completely vanished for x$>$x$_{c}$ (NTI) along with the quenching of the topological surface state. Furthermore, the Peak-A become stronger/weaker by the electron depletion/injection into the Bi$_{2}$Se$_{3}$ in the electrical gating measurement.
The two experimental results provide convincing evidence that Peak-A arises from the 
population of SS, i.e, the optical excitation of electron from 1eV below into into SS. This SS-optical population increases the density of the  surface electron, thus can enhance the topological electrical conduction, which promotes TI device application.  
Similar optically driven SS-population may be realized
in other  TI materials as well, which should be investigated in the future. 
{\it Note added:}  For our $\rm{(Bi_{1-x}In_{x})_{2}Se_{3}}$ films, the bulk transition E$_{op}$ shows a different x-dependent behavior from  Ref.\cite{wu2013sudden}. 
It may come from that carrier doping due to Se-vacancy is sample-dependent for these TI films. See Supplemental Fig.S3\cite{suppley}.
\newline


{\it Acknowledgments.}
This work was supported by the 2016 sabbatical year research grant of the University of Seoul.
JM and SO are  supported by the Gordon and Betty Moore Foundation’s EPiQS Initiative (GBMF4418) and National Science Foundation (NSF) grant EFMA-1542798.

%

\newpage


\begin{figure*}[ht]
\centering
\includegraphics[width=0.7\linewidth]{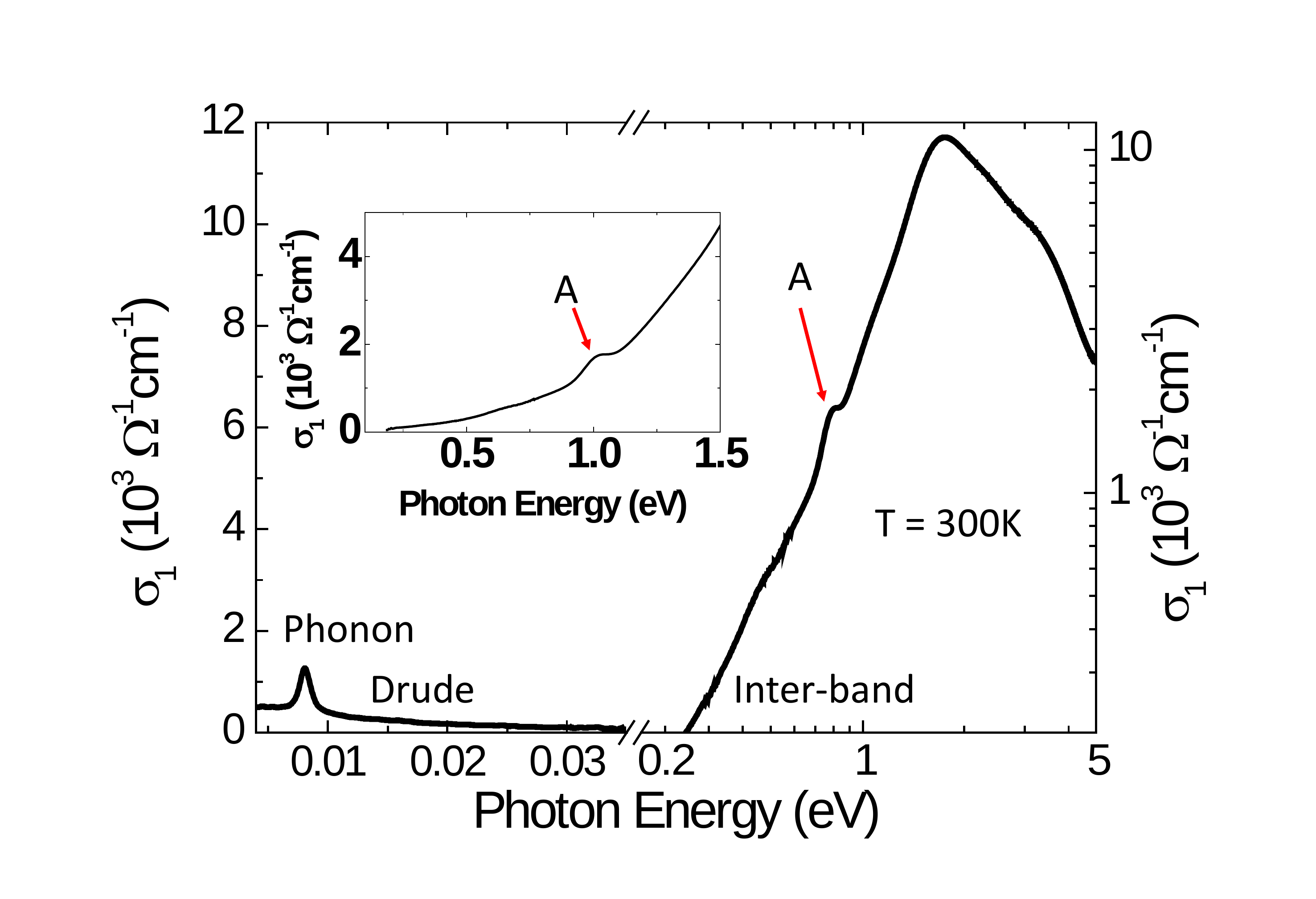}
\caption{
{\bf Optical conductivity of Bi$_{2}$Se$_{3}$ thin film (d=50QL)}
$\sigma_1(\omega)$ shows the Drude, phonon, and interband transition.
Note that log-log scale is used  for the
0.2eV$\leq$$\hbar\omega$$\leq$5eV range.  Inset shows the 
Peak-A in the real scale.    
}
\label{fig1}
\end{figure*}

\begin{figure*}[ht]
\centering
\includegraphics[width=1.0\linewidth]{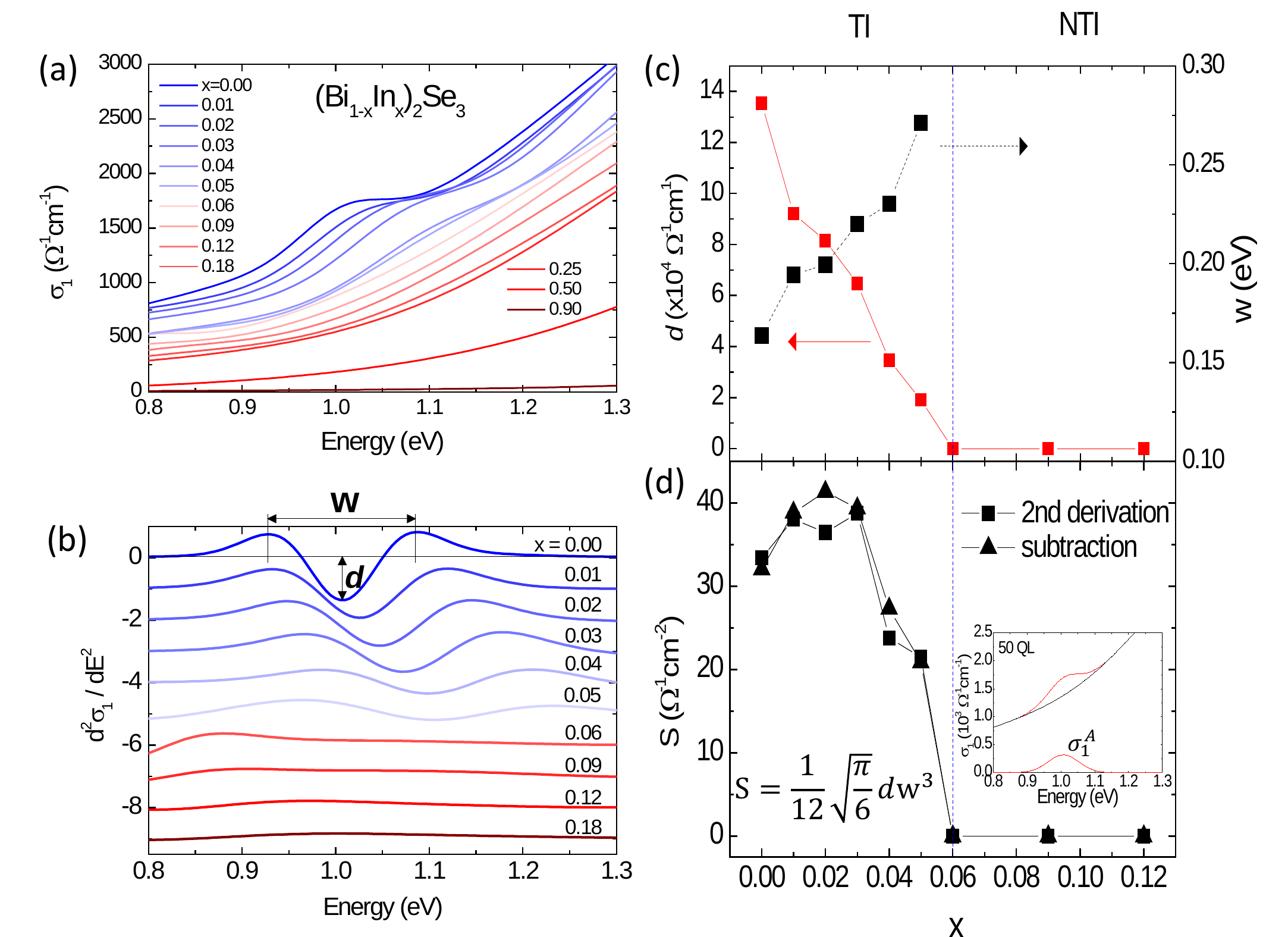}
\caption{
{\bf Evolution of Peak-A in In-substituted  
(Bi$_{1-x}$In$_{x}$)$_{2}$Se$_{3}$ thin films. }
{\bf (a)} 
optical conductivity of (Bi$_{1-x}$In$_{x}$)$_{2}$Se$_{3}$ for the
In-concentration range of 0$\leq$x$\leq$0.9.
{\bf (b)} 
The second derivate of optical conductivity $\frac{d^{2}\sigma_{1}}{d\rm{E}^{2}}$. 
The x-dependent behavior of Peak-A can be traced more clearly in this plot. Here 
w and $d$ denote the distance between the two maxima and  
depth of the dip, respectively.
{\bf (c)} 
The width w and depth $d$ are shown as function of x.
{\bf (d)}
The peak strength S  calculated from S=$\rm{\frac{1}{12}\sqrt{\frac{\pi}{6}}dw^3}$ (see Supplemental Fig.S1\cite{suppley}).
We also calculate S by  S=$\int\sigma_{1}^{A}(\omega)d\omega$ where
$\sigma_{1}^{A}(\omega)$=$\sigma_{1}(\omega)$-$\sigma_{1}^{BG}(\omega)$  ($\sigma_{1}^{BG}$ = polynomial background)
as shown in the inset. 
In (c) and (d), 
the critical concentration x$_{c}$=0.06 of the 
TI to non-TI (NTI) phase transition is highlighted by the vertical line. 
}
\label{fig2}
\end{figure*}

\begin{figure*}[ht]
\centering
\includegraphics[width=1\linewidth]{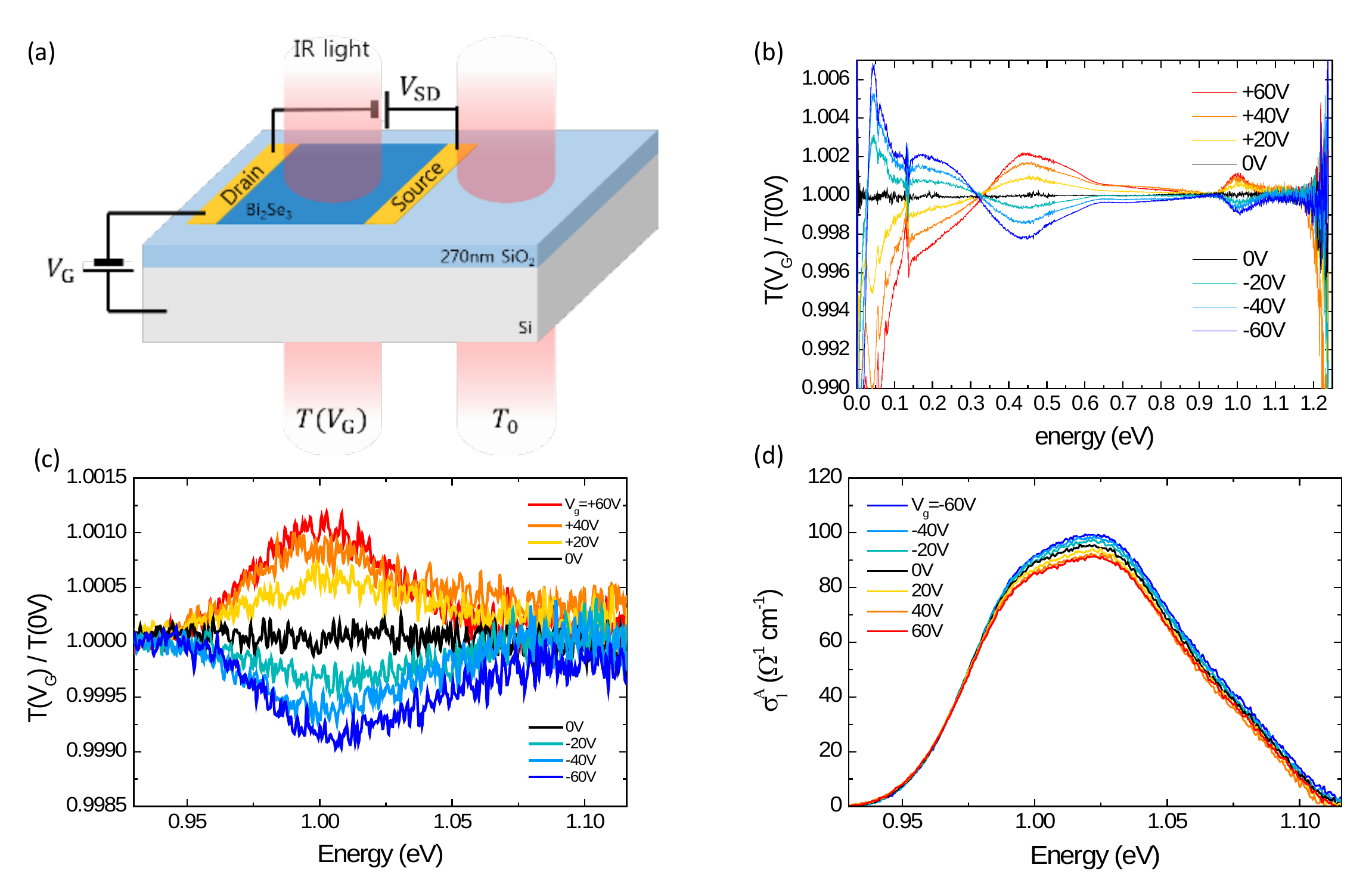}
\caption{
{\bf Gate-driven change of Bi$_{2}$Se$_{3}$.}
{\bf (a)}
Schematic diagram of gate-controlled  transmission measurement of Bi$_{2}$Se$_{3}$ thin film. 
T($V_{\rm{G}}$) is taken with the bias voltage 
$V_{\rm{G}}$ applied between the film and Si.
{\bf (b)} T($V_{\rm{G}}$)/T(0) changes in the Far-IR, mid-IR, and at 1eV.
{\bf (c)} For peak-A T($V_{\rm{G}}$) increases/decreases for the negative/positive $V_{\rm{G}}$, respectively. 
{\bf (d)} 
Optical conductivity of Peak-A, $\sigma_{1}^{A}(\omega)$, was calculated from 
T($V_{\rm{G}}$) in (c) as described in the text. 
}
\label{fig3}
\end{figure*}

\begin{figure*}[ht]
\centering
\includegraphics[width=1.0\linewidth]{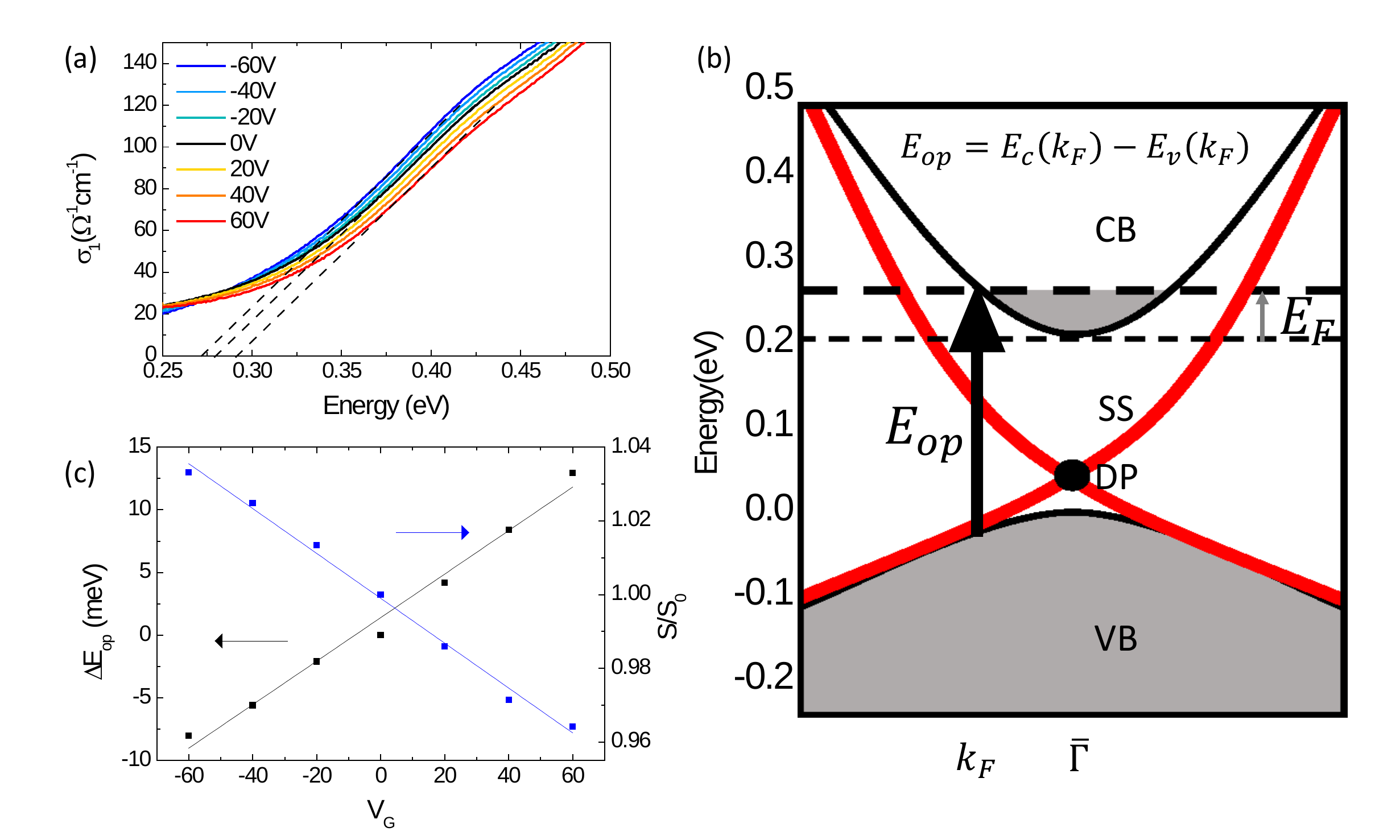}
\caption{
{\bf Estimation of the growth of Peak-A at high gating voltage.}
{\bf (a)}
Gate-driven shift of the inter-band transition.  The onset energy E$_{op}$ is determined by the linear extrapolation of data (dashed line).
{\bf (b)} Schematic diagram of the  VB$\rightarrow$CB interband transition.
The E$_{op}$ corresponds to onset of the VB$\rightarrow$CB. Here 
the Fermi energy E$_{F}$ is measured from the CB bottom. DP stands for Dirac point.
{\bf (c)}  The shift $\Delta$E$_{op}$=E$_{op}(V_{\rm{G}})$-E$_{op}(0)$ and the change of $S$ are 
plotted against $V_{\rm{G}}$. Here
$S$ was calculated by integrating $\sigma_1(\omega)$ in Figure 3c as $S$=$\int\sigma_{1}^{A}(\omega)d\omega$. 
}
\label{fig4}
\end{figure*}

\begin{figure*}[ht]
\centering
\includegraphics[width=0.8\linewidth]{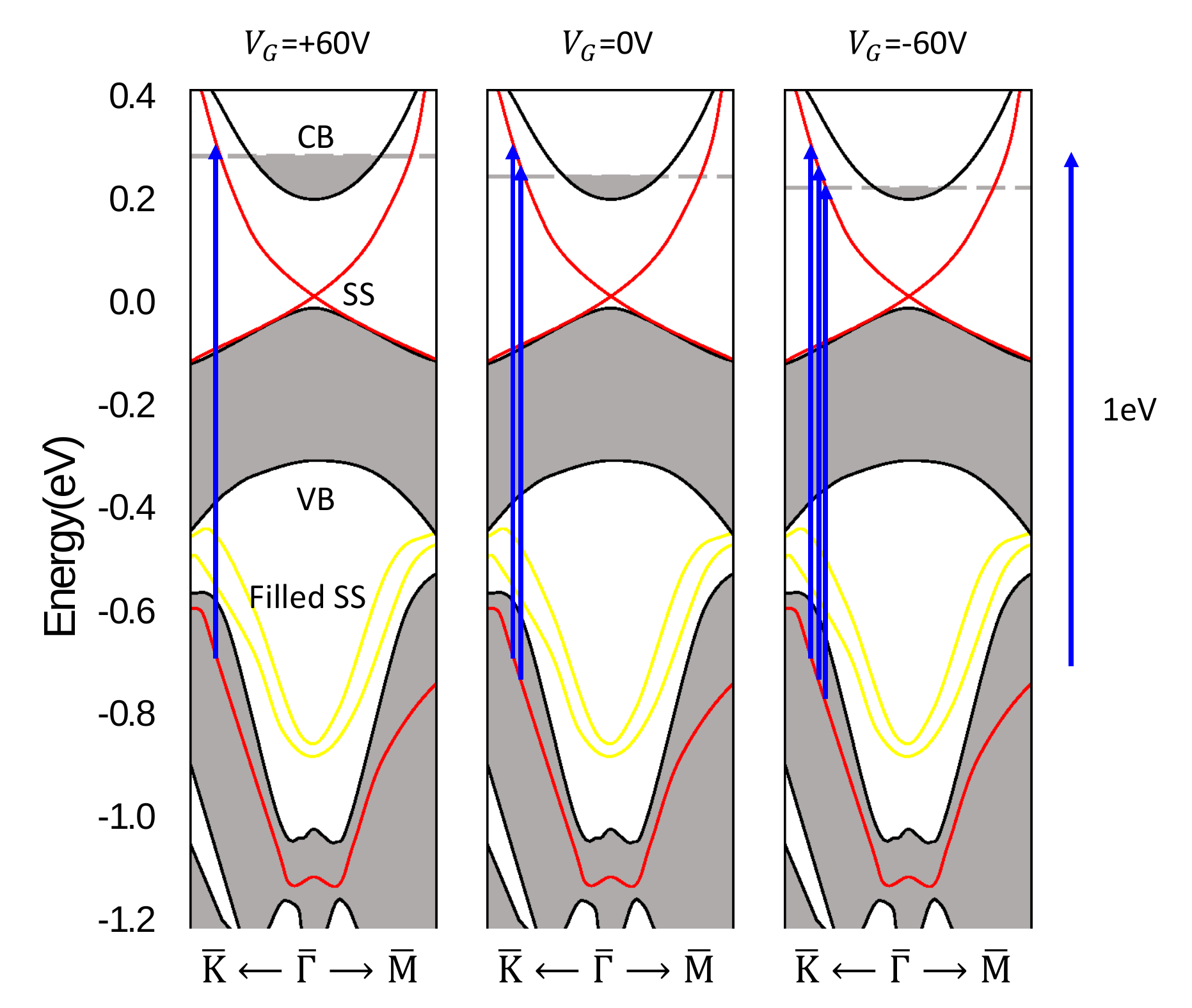}
\caption{
{\bf Schematic band structure and possible origin of Peak-A.}
For $V_{\rm{G}}$=0, electron is excited from the bulk band branch into the empty surface state as highlighted by the blue arrows. For the positive gating, the Fermi level E$_{F}$ shifts up and the 1eV- transition becomes weaker.  For the negative gating E$_{F}$ shifts down and the 1eV- transition becomes stronger. 
The amount of E$_{F}$-shift is exaggerated for clarity.  The  band diagram shown here was redrawn schematically based on experiment \cite{hsieh2009tunable,nechaev2013evidence,hasan2010colloquium,pan2011electronic,wray2010observation,dubroka2017interband,piot2016hole} and  theory \cite{aguilera2013g,guo2016tuning,forster2015ab,hermanowicz2019iodine}.  CB=bulk conduction band, VB=bulk valence band, SS=surface state.
}
\label{Fig5}
\end{figure*}
\newpage

\end{document}